\documentclass[12pt]{article}
\usepackage{amssymb}
\usepackage{graphicx}

\begin{document}

\title{Further evidence that the transition of 4D dynamical
  triangulation is 1st order.}

\author{Bas V. de Bakker\\ Centre for High Energy Astrophysics\\ 
  Kruislaan 403, 1098 SJ Amsterdam, the Netherlands\\ email:
  bas@astro.uva.nl}

\date{March 28, 1996}

\maketitle

\begin{abstract}
  We confirm recent claims that, contrary to what was generally
  believed, the phase transition of the dynamical triangulation model
  of four-dimensional quantum gravity is of first order.  We have
  looked at this at a volume of $64,000$ four-simplices, where the
  evidence in the form of a double peak histogram of the action is
  quite clear.
\end{abstract}

%
\clearpage

\section{Introduction}

Dynamical triangulation is a relatively recent approach to the problem
of formulating a theory of four-dimensional quantum gravity
\cite{We82,AmJu92,AgMi92a}.  Although at the moment only a euclidean
version of the dynamical triangulation model exists, many interesting
results have been obtained in this version.

In dynamical triangulation, the path integral over metrics is replaced
by a sum over ways to glue together a number $N_4$ of four-dimensional
simplices in all possible ways that have a certain fixed topology that
is usually taken to be the sphere $S^4$.  In this procedure, the
continuum euclidean action
\begin{equation}
  S = \frac{1}{16\pi G} \int d^4x \sqrt{g} (2\Lambda - R)
\end{equation}
can be rewritten in the discrete form
\begin{equation}
  S = - \kappa_2 N_2 + \kappa_4 N_4,
  \label{n2action}
\end{equation}
where the $N_i$ are the number of $i$-simplices in the triangulation.
For later reference, we can use the geometrical relation between these
$N_i$
\begin{equation}
  N_0 = \frac{N_2}{2} - N_4 + 2 , 
  \label{nirel}
\end{equation}
to write the discrete action (\ref{n2action}) in terms of the number
of vertices $N_0$
\begin{equation}
  S = - 2 \kappa_2 N_0 + (\kappa_4 - 2 \kappa_2) N_4 + 4 \kappa_2.
  \label{n0action}
\end{equation}

This system turns out to have a phase transition at a critical
$\kappa_2$ value $\kappa_2^c$ that depends somewhat on the volume
$N_4$ and for large volumes converges to a value that was recently
measured as $\kappa_2^c = 1.336(6)$ \cite{AmJu95a}.  Various evidence
is present that indicates that the transition is a continuous one
\cite{AmJu95a,CaKoRe94a}.  Recently, however, data have been presented
\cite{BBKP96} that indicate that the phase transition is a first order
transition.  This point of view was already expressed in
\cite{AgMi92a} where some hysteresis was observed, but this was
retracted in \cite{AgMi92b}.

\section{Simulations}

We have simulated the system at volumes of $32,000$ and $64,000$
simplices at several values of $\kappa_2$ close to the phase
transition.  There is no known method to use ergodic moves that keep
the volume constant and probably no such method can exist
\cite{NaBe93}.  It is known that such a method cannot exist for
manifolds that are unrecognizable and also that some $4$-manifolds
indeed are unrecognizable, although for the $4$-sphere used in our
simulations this is not known.  It would be quite surprising if there
turned out to be a set of local moves whose ergodicity depends on the
topology of the manifold, but for non-local moves such as baby
universe surgery \cite{AmJu95a} this is easier to imagine.

To make sure the moves in the simulations are ergodic, we therefore
have to allow fluctuations of the number of simplices $N_4$.  This is
done by the usual method of allowing the number of simplices to vary,
but at the same time adding a quadratic term to the action to keep
this number close to some desired value $V$.  The action then becomes
\begin{equation}
  S = -\kappa_2 N_2 + \kappa_4 N_4 + \gamma (N_4 - V)^2 ,
  \label{modaction}
\end{equation}
where $\gamma$ is a parameter that controls the volume fluctuations
and that we set to $5 \cdot 10^{-4}$.  Because a constant in the
action is irrelevant, it is possible to eliminate either $\kappa_4$ or
$V$ from the action, but we will not do so.  In the simulations, the
magnitude of the fluctuations in $N_4$ is
\begin{equation}
  \delta N_4 = \sqrt{\langle N_4^2 \rangle - \langle N_4 \rangle^2} =
  \sqrt{\frac{1}{2\gamma}} \approx 30.
\end{equation}

Note also that the modification of the action in (\ref{modaction})
only depends on $N_4$.  Therefore, the relative weights of
configurations at a particular value of $N_4$ does not change with
respect to the original action.

The fluctuations of $N_4$ introduce some extra fluctuations in the
values of $N_0$ that we are going to measure.  Because our parameter
$\kappa_2$ couples to the number of triangles $N_2$, one would at
first sight consider measuring $N_2$.  The reason we use $N_0$ is that
it suffers much less from these extra fluctuations.  This can be seen
as follows.  Because of the relation (\ref{nirel}), the ratio of the
fluctuations $\delta N_2 / \delta N_0$ at fixed $N_4$ equals $2$.
However, the ratio of the $N_i$ fluctuations due to the volume
fluctuations at fixed energy density (that is fixed $N_i/N_4$) is
$\delta N_2 / \delta N_0 = N_2 / N_0 \geqslant 10$, the exact ratio
depending on $\kappa_2$.

\section{Results}

\begin{figure}[t]
  \includegraphics[angle=-90,width=\textwidth]{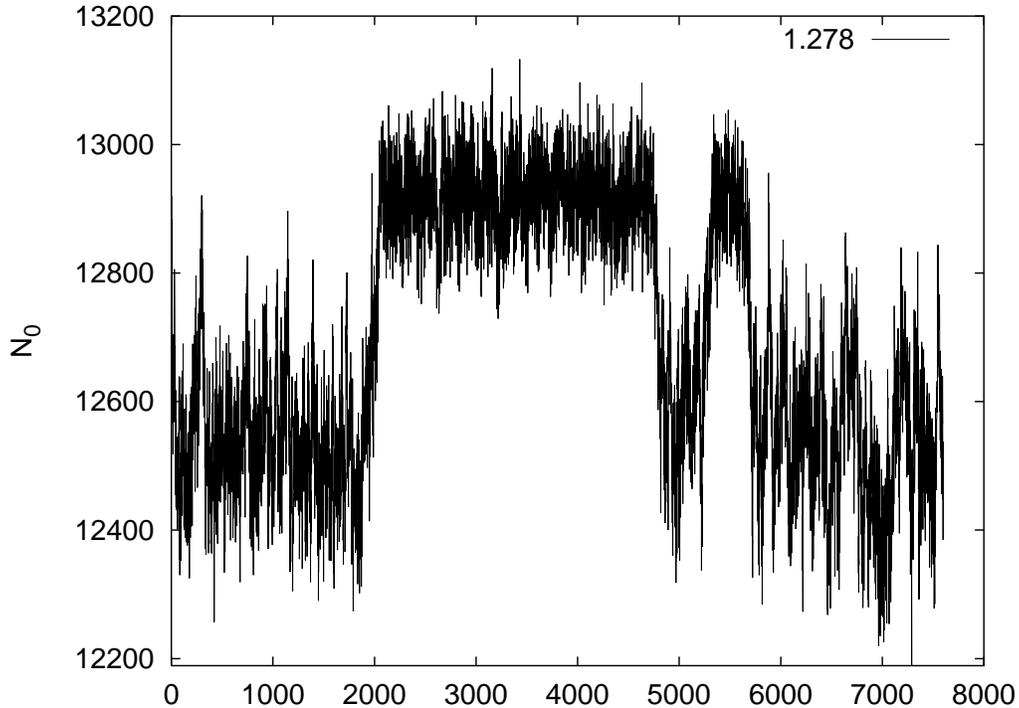}
  \caption{Computer time history of $N_0$, horizontal units are $100$
    sweeps, $\kappa_2 = 1.278$.}
  \label{n0_6fig}
\end{figure}

\begin{figure}[t]
  \includegraphics[angle=-90,width=\textwidth]{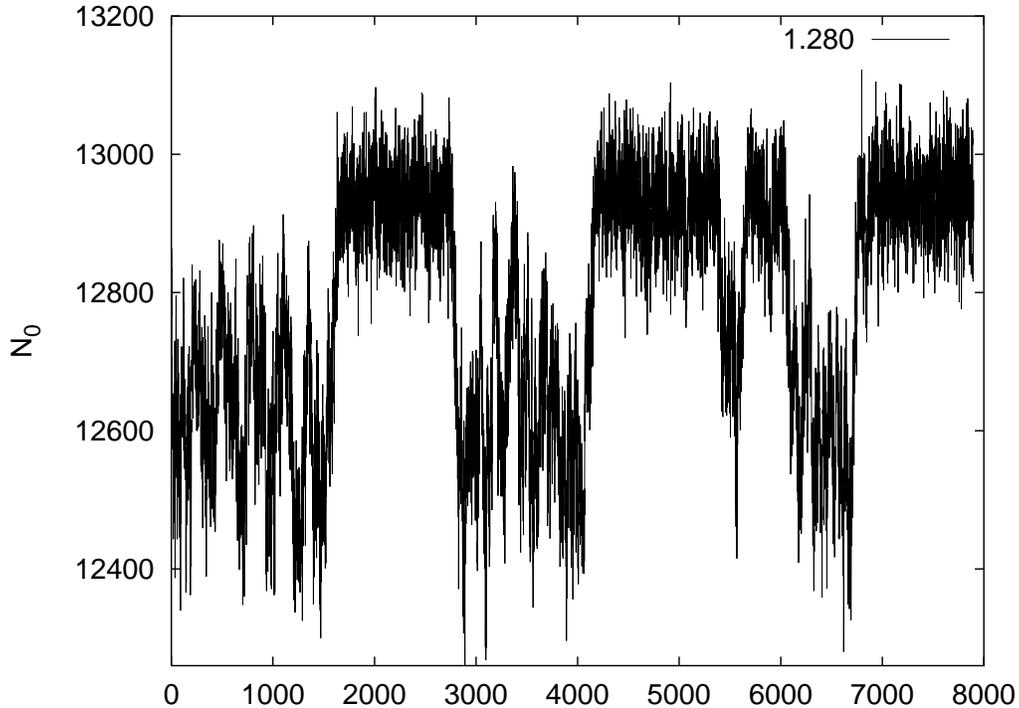}
  \caption{Computer time history of $N_0$, horizontal units are $100$
    sweeps, $\kappa_2 = 1.280$.}
  \label{n0_8fig}
\end{figure}

Two time histories of $N_0$ are plotted in figures~\ref{n0_6fig}
and~\ref{n0_8fig}.  The horizontal units are $100$ sweeps, where we
define a sweep as $N_4$ accepted moves.  The quantity $N_0$ (the
number of vertices in the configuration) is at fixed $N_4$ and up to a
constant directly proportional to the action.  We can see this from
equation (\ref{n0action}).  Several other time histories that were
made show the same effect: there are two states, one at high $N_0$ and
the other at low $N_0$.  The system stays in one of these states for a
long time and occasionally the system flips from one state to the
other.  This is a good indication that there are two separate minima
in the free energy, creating a first order phase transition.

\begin{figure}[t]
  \includegraphics[angle=-90,width=\textwidth]{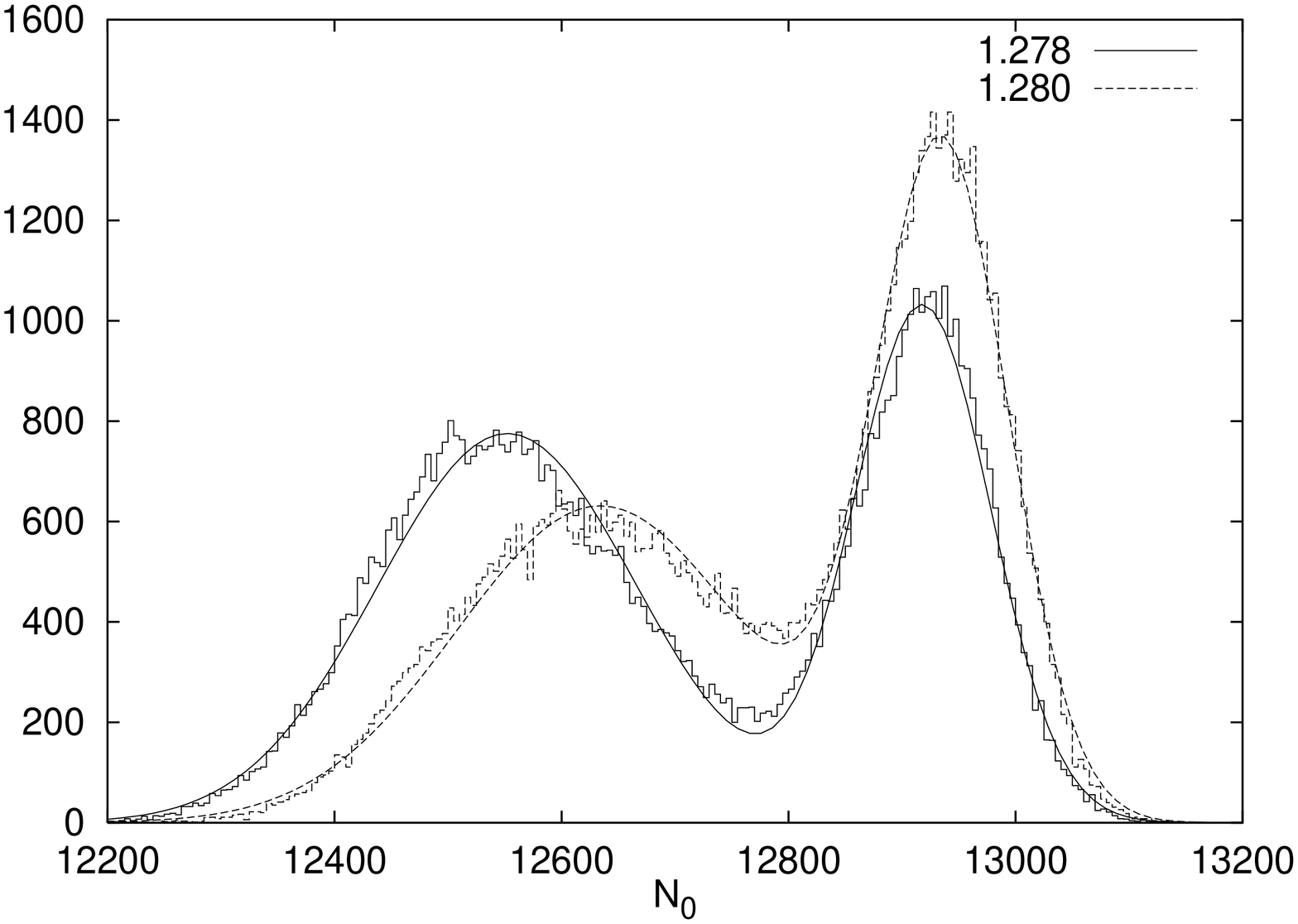}
  \caption{Histogram of $N_0$ values at $N_4 = 64,000$.}
  \label{histn0fig}
\end{figure}

A histogram of observed $N_0$ values is plotted in
figure~\ref{histn0fig}.  The size of the bins is $5$, while $N_0$
values were taken each $10$ sweeps.  Because of the fluctuations in
$N_4$ explained above, the $N_0$ values are slightly more spread out
than what would be the case if $N_4$ were really fixed.  As the
fluctuations in $N_4$ are about $30$ and $N_0/N_4 \approx 1/6$, these
extra fluctuations in $N_0$ are about $5$, that is the size of one
bin.

The double peak structure in figure~\ref{histn0fig} is clear.  It
should be obvious from the limited time histories in
figures~\ref{n0_6fig} and~\ref{n0_8fig} that the relative strenghts of
the peaks in the histogram are not very significant.  To find the
actual relative contribution of the two kinds of configurations to the
partition function would require many more flips between the states,
which would take inordinate amounts of computer time.

The fits shown in the figure are fits to a double gaussian
\begin{equation}
  C_1 \exp \left( \frac{-(N_0 - \mu_1)^2}{2 \sigma_1} \right) +
  C_2 \exp \left( \frac{-(N_0 - \mu_2)^2}{2 \sigma_2} \right).
\end{equation}

\begin{figure}[t]
  \includegraphics[angle=-90,width=\textwidth]{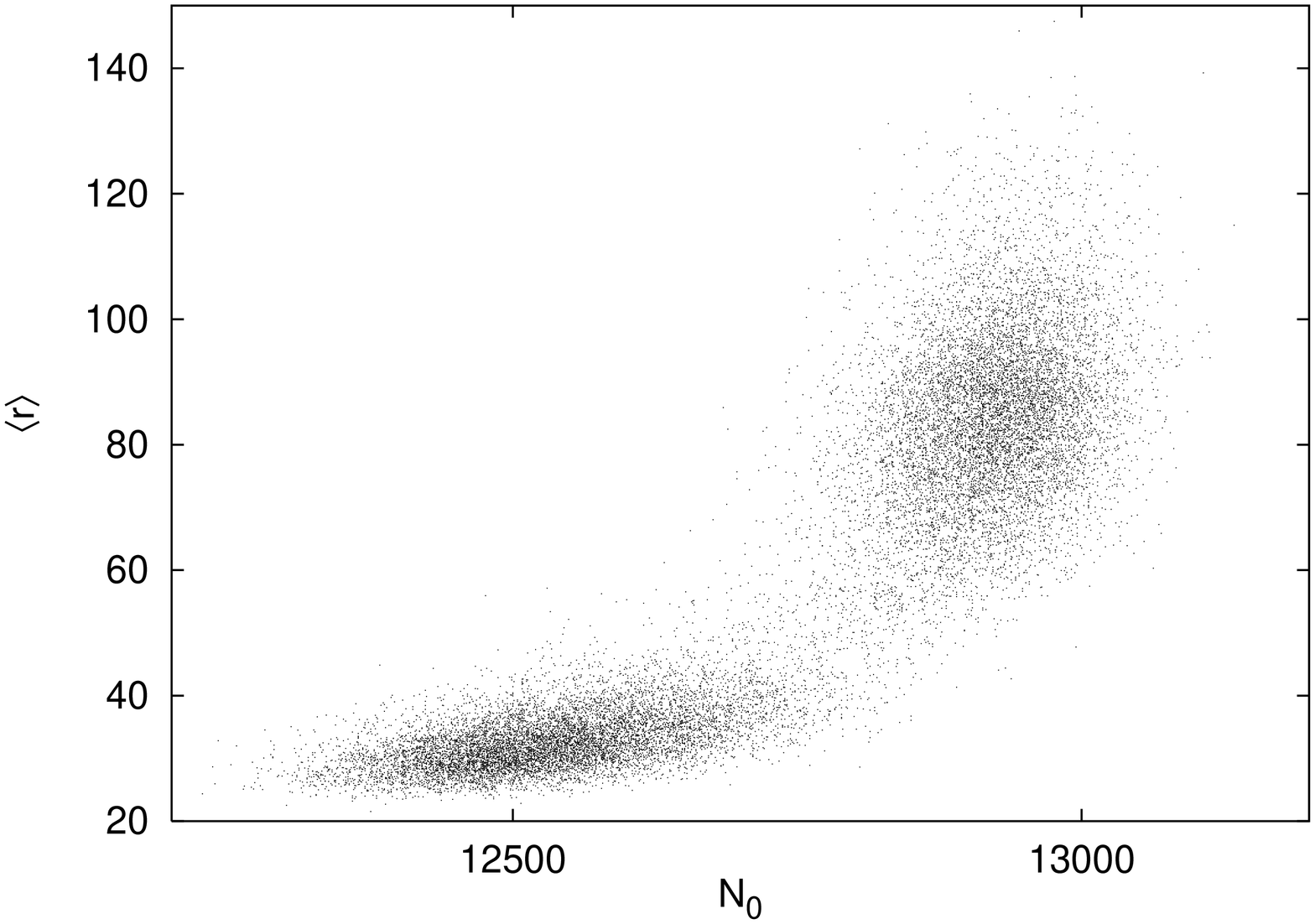}
  \caption{Correlation of $N_0$ with the average distance, $\kappa_2 =
    1.278$.}
  \label{dist6fig}
\end{figure}

The two states with large and small $N_0$ are directly related to the
crumpled or elongated structure of the configuration.  This structure
is quantified using the average distance between any two simplices in
a particular configuration.  Figure~\ref{dist6fig} shows the
correlation of the two quantities.  There seem to be two separated
areas, one where $N_0$ and the average distance are small, and one
where both are large.

Because we fix $\kappa_4$ in the action (\ref{modaction}) at some
reasonable but arbitrarily chosen value in our simulations, the actual
average volume is not exactly the $64,000$ we mention, but $63,912$.
This also means that the transition occurs at a slightly lower value
of $\kappa_2$ than if the average volume were exactly $64,000$.

\begin{figure}[t]
  \includegraphics[angle=-90,width=\textwidth]{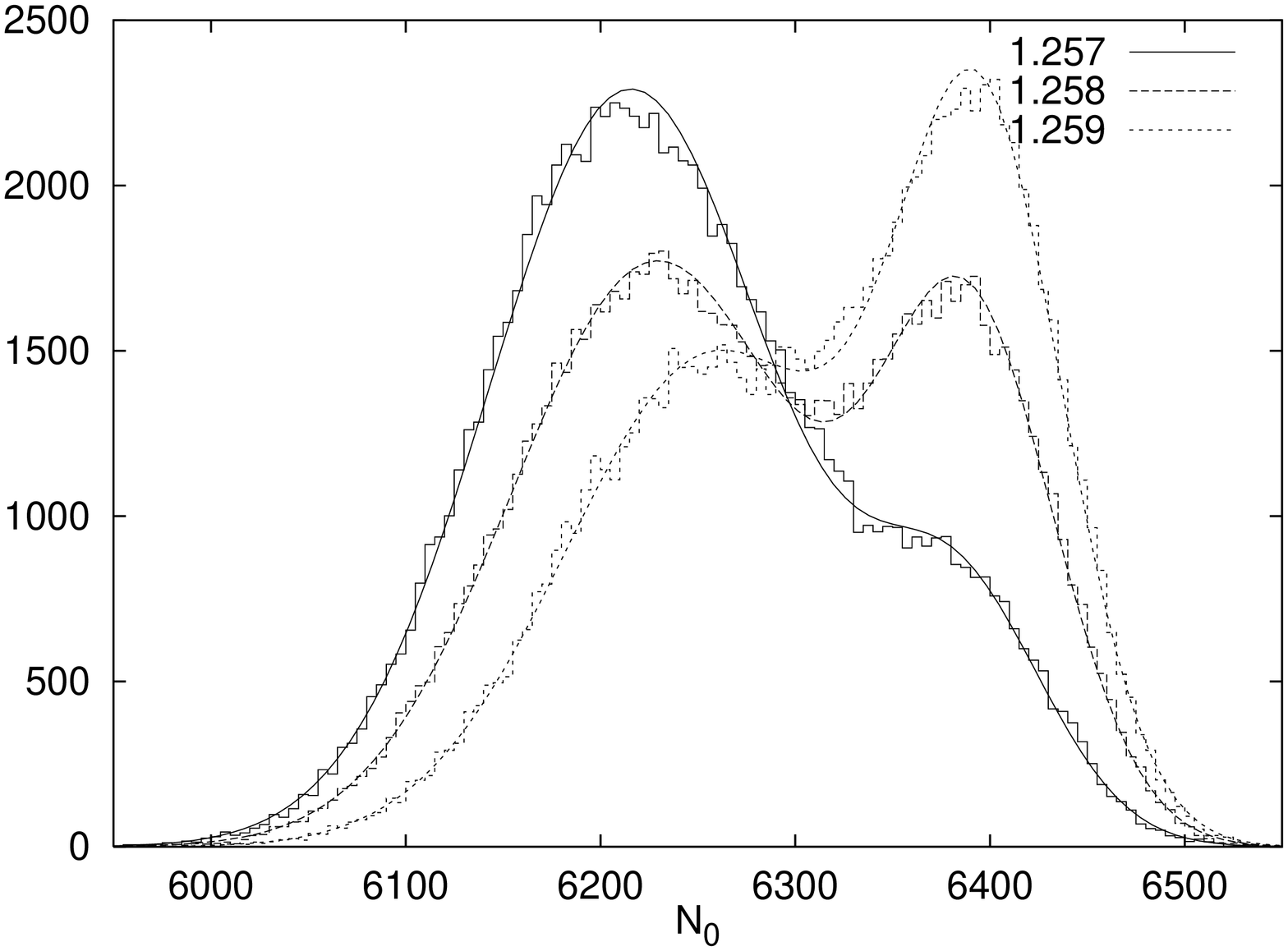}
  \caption{Histogram of $N_0$ values at $N_4 = 32,000$.}
  \label{n0_32fig}
\end{figure}

For comparison, we have also plotted a histogram of $N_0$ values at
the smaller volume of $N_4 = 32,000$.  In this case the actual average
volume was $N_4 = 31,911$.  Only at $\kappa_2 = 1.258$ can we see a
double peak structure, and even that one is quite weak.  This shows
that the effect grows with the volume.  Also, the distance between the
peaks grows roughly linearly with the volume, from $160$ at $N_4 =
32,000$ and $\kappa_2 = 1.258$ to $366$ and $301$ at the two
$\kappa_2$ values shown at $N_4 = 64,000$.  In other words, if we take
an intensive quantity like $N_0/N_4$ the distance between the peaks
stays constant.  If the transition was second order and the double
peak structure a finite volume effect, this distance would shrink with
the volume.

\section{Finite size scaling}

\begin{figure}[t]
  \includegraphics[angle=-90,width=\textwidth]{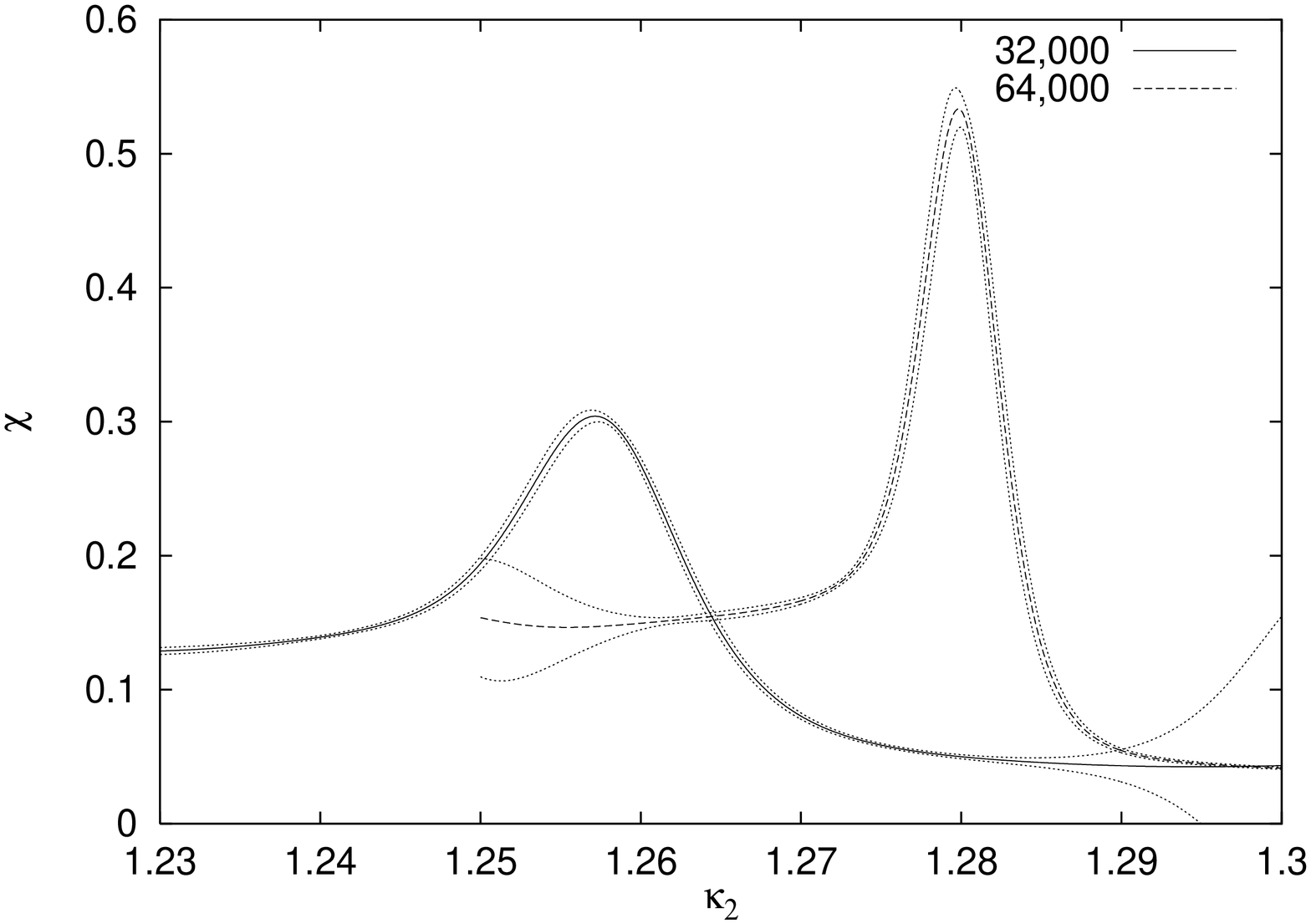}
  \caption{Vertex susceptibility as a function of $\kappa_2$ at the
    two volumes used.}
  \label{n0susfig}
\end{figure}

Using the Monte Carlo runs we made at various values of $\kappa_2$, we
can calculate the susceptibility
\begin{equation}
  \chi = \frac{1}{N_4} \left( \langle N_0^2 \rangle - \langle N_0
  \rangle^2 \right) ,
\end{equation}
as a function of $\kappa_2$.  To cover intermediate values of
$\kappa_2$, we use the Ferrenberg-Swendsen reweighting procedure
\cite{FeSw88,FeSw89}.  The results are plotted in
figure~\ref{n0susfig}.  The errors (represented by the dotted curves)
were generated using the jackknife method.  The large errors at the
end of the curves arise because the reweighting procedure cannot
extrapolate well far from the actual $\kappa_2$ values that were used
in the simulations.  The maxima occur at $\kappa_2 = 1.257(1)$ and
$\kappa_2 = 1.280(1)$ respectively.

The heights of the susceptibility peaks allow us to calculate some of
the finite size scaling exponents of the phase transition.  Some
caution should be exercised when interpreting the results, because
only two volumes were available.  Simulating to the same accuracy at
another large volume would take too long, while smaller volumes showed
no sign of a first order transition and therefore are apparently too
small to give meaningful large volume behaviour.

We will first look at the susceptibility exponent $\Delta$ defined by
\begin{equation}
  \chi_\mathrm{max} \propto N_4^{\Delta} .
\end{equation}
In a regular spin system with a second order transition, $\Delta$
would be related to the susceptibility exponent $\gamma$, the
correlation length exponent $\nu$ and the dimension $d$ by the
relation $\Delta = \gamma/\nu d$, but whether such a relation also
holds in simplicial quantum gravity is not clear.  In particular, one
may wonder what the dimension of the system is.  In \cite{BaSm95a} a
scaling dimension $d_s$ was defined that turned out to be compatible
with $4$, but we will not attempt to use that in the present
discussion.

The susceptibility exponent $\Delta$ following from the data of
figure~\ref{n0susfig} is $0.81(4)$.  This is not compatible with an
exponent of $1$ expected at a first order phase transition.  It is
however much larger than the value of $0.259(7)$ obtained at volumes
up to $8,000$ in \cite{CaKoRe94a}.  Apparently the number goes up with
the volume and may very well approach $1$ at even larger volumes.  One
should also note that the number of $0.81(4)$ was obtained by taking
the absolute values of the peaks, and not their height relative to the
``background'' susceptibility (arising from the fluctuations within
one free energy minimum) that does not increase with the volume.
Doing this would result in a higher exponent, but we cannot simply
substract the value observed away from the transition, because the
background susceptibility is so different in both phases.

Another exponent to consider is the one that describes the width of
the susceptibility peak
\begin{equation}
  \delta \kappa_2 \propto N_4^{-\Gamma} .
\end{equation}
As the width of the peak, we arbitrarily take the width at $75\%$ of
the peak height.  Taking the usual definition of the width at half the
maximum height would take us too far into the left hand tail of the
curve at $N_4 = 32,000$.  The exponent $\Gamma$ we get from the data
in figure~\ref{n0susfig} is $1.24(18)$.  This is compatible with the
value of $1$ expected at a first order transition.  In this case, the
background susceptibility makes the exponent come out higher than its
actual value in the infinite volume limit.

The finite size scaling exponent that governs the change of the
apparent critical value $\kappa_2^c(N_4)$ with the volume cannot be
determined from only two volumes.

\section{Discussion}

We have seen that at a volume of $N_4 = 64,000$ the time history
flipping and double peak structure in the action are very clear
indicating a first order transition.  We calculated some finite size
scaling exponents, but due to the small number of volumes they cannot
be interpreted as the definitive values.

It has been suggested \cite{BBKP96} that constraining the volume in
the simulations may create an artificial potential barrier, causing
the flipping between the elongated and the crumpled state.  This seems
unlikely to me, because of the following.  There cannot be an energy
barrier between the states, because intermediate states by definition
have intermediate energy.  And a barrier in the volume, in the sense
that one needs a higher volume to get from one state to the other,
simply means that at the value of $N_4$ under consideration the number
of intermediate states is small, meaning that the first order
transition is genuine.

If the process is ergodic in practice, all states at fixed $N_4$ will
be visited with frequency proportional to their respective Boltzmann
weights.  In this case there is obviously no problem and the double
peak structure is real.  If the double peak structure is caused by a
non-ergodicity of the moves, this can only be because many of the
\emph{intermediate} states (at the gap between the peaks) do exist,
but are unreachable.  This seems unlikely, because such unreachable
states have not been observed \cite{AmJu94b,Ba95b} and if they exist
in significant numbers one would expect them to have some extremal
feature and not be in the middle of the $N_2$ distribution.

As has been pointed out earlier \cite{BaSm95a}, the system may scale
even away from the phase transition, making a continuous transition
not vital for continuum behaviour of the theory.  As an example, in
the case of $\mathbb{Z}_N$ gauge theory \cite{FrSp82,Al83,AlBo83} the
system has infinite correlation length in a whole region of values of
the coupling constant (the Coulomb phase).  A different scenario can
be found in SU(2) gauge-Higgs theory (see e.g.\ \cite{Ya95,Ka95,Ja95}
and references therein).  This system has a first order transition,
but decreasing the lattice spacing increases the correlation length in
lattice units in such a way that the system still scales.

\section*{Acknowledgements}

The author would like to thank J.~Smit and P.~Bia{\l}as for useful
discussions.  The simulations were carried out on the Parsytec
PowerXplorer and CC/40 at IC3A and the IBM SP1 and SP2 at SARA.


\begin{thebibliography}{99}

\bibitem{We82} D.~Weingarten, Nucl. Phys. B210 [FS6] (1982) 229.
  
\bibitem{AmJu92} J.~Ambj{\o}rn and J.~Jurkiewicz, Phys. Lett. B278
  (1992) 42.

\bibitem{AgMi92a} M.E.~Agishtein and A.A.~Migdal, Mod. Phys. Lett.  A7
  (1992) 1039.
  
\bibitem{AmJu95a} J.~Ambj{\o}rn and J.~Jurkiewicz, Nucl. Phys. B 451
  (1995) 643.
  
\bibitem{CaKoRe94a} S.~Catterall, J.~Kogut and R.~Renken, Phys. Lett.
  B 328 (1994) 277.
  
\bibitem{BBKP96} P.~Bia{\l}as, Z.~Burda, A.~Krzywicki and
  B.~Petersson, \textsl{Focusing on the fixed point of 4D simplicial
    gravity}, preprint LPTHE Orsay 96/08, BI-TP 96/05, ITFA-96-2,
  hep-lat/9601024.
  
\bibitem{AgMi92b} M.E.~Agishtein and A.A.~Migdal, Nucl. Phys. B 385
  (1992) 395.
  
\bibitem{NaBe93} A.~Nabutovsky and R.~Ben-Av, Commun. Math. Phys. 157
  (1993) 93.

\bibitem{FeSw88} A.M.~Ferrenberg and R.H.~Swendsen, Phys. Rev. Lett.
  61 (1988) 2635.
  
\bibitem{FeSw89} A.M.~Ferrenberg and R.H.~Swendsen, Phys. Rev. Lett.
  63 (1989) 1195.

\bibitem{BaSm95a} B.V.~de Bakker and J.~Smit, Nucl. Phys. B 439 (1995)
  239.

\bibitem{AmJu94b} J.~Ambj{\o}rn and J.~Jurkiewicz, Phys. Lett. B 345
  (1995) 435.

\bibitem{Ba95b} B.V.~de Bakker, Phys. Lett. B 348 (1995) 35.
  
\bibitem{FrSp82} J.~Fr\"olich and T.~Spencer, Commun. Math. Phys. 83
  (1982) 411.

\bibitem{Al83} V.~Alessandrini, Nucl. Phys. B 215 (1983) 337.

\bibitem{AlBo83} V.~Alessandrini and Ph.~Boucaud, Nucl. Phys. B 225
  (1983) 303.
  
\bibitem{Ya95} L.G.~Yaffe, \textsl{The electroweak phase transition: a
    status report}, preprint hep-ph/9512265.
  
\bibitem{Ka95} K.~Kajantie, Nucl. Phys. B (Proc. Suppl.) 42 (1995)
  103.
  
\bibitem{Ja95} K.~Jansen, \textsl{Status of the finite temperature
    electroweak phase transition on the lattice}, preprint
  DESY-95-169, hep-lat/9509018.

\end{thebibliography}
\end{document}